\documentclass[letter]{aa}  

\usepackage{graphicx}
\usepackage{txfonts}
\usepackage{lipsum}
\usepackage{subcaption}         
\usepackage{lscape}             
\usepackage{placeins}           
 \usepackage{multirow}                               

\begin{document}







\title{Milliarcsecond-scale spectrum of the persistent radio source associated with FRB\,20190417A and constraints for FRB\,20181030A}




\titlerunning{Milliarcsecond spectrum of the PRS associated with FRB 20190417A}

%

   \author{G. Bruni\inst{1}
        \and L. Piro\inst{1}
        \and Y.-P. Yang\inst{2}
        \and L. Nicastro\inst{3}
        \and A. Rossi\inst{3}
        \and E. Palazzi\inst{3} 
        \and E. Maiorano\inst{3}
        \and S. Savaglio\inst{4,3,5}
        \and B. Zhang\inst{6,7,8,9}
        }

   \institute{INAF -- Istituto di Astrofisica e Planetologia Spaziali, via Fosso del Cavaliere 100, 00133 Rome, Italy\\
             \email{gabriele.bruni@inaf.it}
             \and
            South-Western Institute for Astronomy Research, Yunnan University, Kunming, China. 
            \and
            INAF -- Osservatorio di Astrofisica e Scienza dello Spazio di Bologna, via Piero Gobetti 93/3, I-40129 Bologna, Italy
            \and
            Dipartimento di Fisica, Università della Calabria, Arcavacata di Rende, Italy 
            \and
            Laboratori Nazionali di Frascati, INFN (Istituto Nazionale di Fisica Nucleare), Frascati, Italy 
            \and
            The Hong Kong Institute for Astronomy and Astrophysics, the University of Hong Kong, Pokfulam Road, Hong Kong
            \and
            Department of Physics, the University of Hong Kong, Pokfulam Road, Hong Kong
            \and
            Nevada Center for Astrophysics, University of Nevada, Las Vegas, NV, USA. 
            \and
            Department of Physics and Astronomy, University of Nevada, Las Vegas, NV, USA
            }

   \date{Received September 30, 20XX}
\abstract
{}
{We confirm the compact nature and constrain the radio spectra of candidate persistent radio sources (PRSs) associated with repeating fast radio bursts (FRBs), and we increase the number of PRSs with spectral indices measured using very long baseline interferometry (VLBI).}
{We performed observations with the European VLBI Network (EVN) at 5 and 8 GHz targeting two candidate PRSs that were identified in a recent VLA survey of repeating FRBs. We measured the flux densities and upper limits at milliarcsecond resolution and combined them with published data at lower frequencies to derive spectral constraints.}
{We detect a compact source associated with FRB\,20190417A at 5\,GHz with a flux density of $150\pm45\,\mu$Jy, but no detection is obtained at 8\,GHz. The source is unresolved on milliarcsecond scales, implying a projected physical size $\lesssim15$\,pc and a brightness temperature $T_{\rm b} > 3.8\times10^{5}$\,K, confirming its non-thermal nature. The combination of our measurement with the 1.4\,GHz VLBI data yields a spectral index $\alpha=-0.19\pm0.29$, consistent with a nearly flat spectrum. This makes FRB\,20190417A only the second PRS with a spectral index constrained using VLBI data. The inferred luminosity places the source on the proposed $L_{\nu}$--$|\rm RM|$ relation. When this source is included, the scatter is $\sigma_{\Delta}=0.67$, corresponding to $\hat{\alpha}|\epsilon|\sim1.5$. This value is roughly consistent with scenarios involving forward shocks in the free-expansion phase or young pulsar wind nebulae. For the candidate PRS associated with FRB\,20181030A, we report $5\sigma$ upper limits of $80\,\mu$Jy at 5\,GHz and $150\,\mu$Jy at 8\,GHz, corresponding to $L_{5\,\rm GHz}\lesssim3.8\times10^{25}$\,erg\,s$^{-1}$\,Hz$^{-1}$, and implying a steep spectral index ($\alpha\lesssim-1.2$) if the VLA emission arises from a compact component.}
{Our results highlight the importance of VLBI in isolating compact emission from FRB engines, and they provide one of the few spectral constraints for PRSs at milliarcsecond resolution. The consistency of FRB\,20190417A with the $L_{\nu}$--$|\rm RM|$ relation supports a nebular origin for the persistent emission.}


   \maketitle
   \nolinenumbers
\section{Introduction}

Fast radio bursts (FRBs) are bright millisecond-duration sources of extragalactic origin that have only been observed at radio wavelengths so far. The vast majority are singular events, and only a small fraction ($\sim$2.5\%) are currently observed to repeat \citep{Abbott_2026}. It is plausible that repetition is a common property that is often below current detection thresholds \citep{Kirsten+24}. Despite the large number of detections and extensive multi-wavelength follow-up campaigns, counterparts at other wavelengths remain elusive, with the notable exception of a Galactic magnetar \citep{Bochenek20}. This observational landscape still accommodates a wide range of progenitor models.

Magnetars can reproduce many of the observed properties of FRBs \citep{Zhang20}, and they are among the leading candidates for their central engines. However, magnetars themselves can arise from different evolutionary channels, including core-collapse supernovae and compact binary mergers \citep{Margalit19, niu22}. FRB\,20200120E, located in a globular cluster \citep{Kirsten22}, provides a possible example of the compact binary merger channel. These different formation pathways are expected to imprint distinct signatures on the host galaxies and local environments, but the current sample remains too limited to draw firm conclusions.

Significant progress has been achieved by probing FRB environments across multiple wavelengths and spatial scales. High-resolution radio interferometry, combined with optical/IR spectroscopy and X-ray observations, has enabled the identification and characterisation of host galaxies and star-forming regions down to sub-arcsecond scales \citep{2017ApJ...834L...7T,Bhandari20,2021A&A...656L..15P,Bhandari22,Bruni2024Nat}. In a few remarkable cases, these efforts have also revealed compact persistent radio sources (PRSs) that are spatially coincident with the FRB position, such as in FRB\,20121102A \citep{chatterjee17,marcote17}, FRB\,20190520B \citep{niu22}, FRB\,20201124A \citep{Bruni2024Nat}, and more recently, FRB\,20240114A \citep{Bruni2025AandA}. In addition, VLBI follow-up observations have recently confirmed the compact nature of one of the candidate PRSs identified by \citet{Ibik2024}, further supporting its association with the FRB\,20190417A \citep{2026ApJ...996L..16M}.

The presence of a compact PRS provides a unique probe of the immediate environment of the FRB central engine. In the magnetar scenario, this emission can arise from a magnetised nebula powered by continuous energy injection \citep{Murase16,margalit18}. The large Faraday rotation measures (|RM|) observed in some FRBs indicate dense and highly magnetised surroundings, motivating a relation between the radio luminosity of the PRS and the |RM| \citep{yang20,yang22}. This relation has been reinforced by the discovery of the PRS associated with FRB\,20201124A, which extended the explored parameter space by some orders of magnitude \citep{Bruni2024Nat}.


A recent systematic survey with the Karl G. Jansky Very Large Array (VLA) of 37 repeating fast radio bursts (FRBs) discovered by the Canadian Hydrogen Intensity Mapping Experiment (CHIME) has identified two candidate PRSs consistent with the FRB positions \citep{Ibik2024}. The two sources lie in the region of the radio luminosity versus |RM| plane expected for magnetised nebulae, providing further support to the proposed correlation. However, due to the limited angular resolution of the VLA observations, contamination from star formation within the host galaxies cannot be excluded, leaving the nature of these sources uncertain.

Observations with very long baseline interferometry (VLBI) are crucial to confirm the compactness and association of these candidates with the FRB engine. Recently, one of the candidates reported by \citet{Ibik2024} has been followed up with the European VLBI Network (EVN), leading to the detection of a compact source at milliarcsecond scales \citep{2026ApJ...996L..16M}. This result strongly supports the identification of this source as a genuine PRS and highlights the key role of VLBI in isolating the nuclear component from host galaxy emission. Further details on the optical photometric and spectroscopic properties of the host galaxies of FRB\,20190417A and FRB\,20181030A are presented in \citet{2026ApJ...996L..16M} and \citet{Bhardwaj_2021}.

However, robust spectral information for PRSs at milliarcsecond resolution remains extremely scarce. To date, only the radio spectrum of the PRS associated with FRB\,20121102A is constrained using VLBI data. Recent VLBI detections of candidate PRSs \citep{2026ApJ...996L..16M} have confirmed their compact nature, but lack the multi-frequency coverage required to derive a reliable spectral index.

In this Letter, we present EVN follow-up observations at 5 and 8 GHz of the two candidate persistent radio sources reported by \citet{Ibik2024}, 20190417A-S1 and 20181030A-S1, with the goal of confirming their compact nature and constraining their radio spectra.
We adopted a flat $\Lambda$CDM cosmology with 
$H_0$ = 67.36 km s$^{-1}$ Mpc$^{-1}$, 
$\Omega_{\rm m}$ = 0.315, and 
$\Omega_{\Lambda}$ = 0.685 \citep{2020A&A...641A...6P}.


\section{Results}

Details on observations, on the detection of 20190417A-S1, and on the non-detection of 20181030A-S1 and its implications, are given in the appendices \ref{appendix:detection} and \ref{appendix:nondetection}, respectively. Here, we focus on the spectral properties of 20190417A-S1.


We compared our EVN integrated flux density at 5\,GHz ($S_{5\,\mathrm{GHz}} = 150 \pm 45\,\mu$Jy) with the flux density at 1.4\,GHz reported by \citet[$S_{1.4\,\mathrm{GHz}} = 191 \pm 39\,\mu$Jy]{2026ApJ...996L..16M}. Assuming a power-law spectrum $S_{\nu} \propto \nu^{\alpha}$, we obtained $\alpha_{5/1.4\,\mathrm{GHz}} = -0.19 \pm 0.29$, where the uncertainty was derived by propagating the fractional flux errors. This value indicates that the spectrum is approximately flat or only mildly declined between 1.4 and 5\,GHz, without evidence of a steep spectrum.

Since the two measurements were obtained at different epochs, 
intrinsic variability cannot be ruled out. Importantly, both probe 
the compact PRS emission at milliarcsecond scales with negligible 
contamination from host-galaxy star formation. A recent reprocessing 
of the same VLA 1.52\,GHz observations originally presented by 
\citet{Ibik2024} by \citet{2026A&A...709L...1B} yielded 
$S_{1.52\,\mathrm{GHz}} \approx 250\,\mu$Jy after full primary-beam 
correction. This is higher by a factor of $\sim$1.3 than the original 
VLA value and the VLBI measurement. The resulting discrepance is 
$\sim$1.5$\sigma$, suggesting a possible contribution from extended emission at arcsecond scales, although at a moderate statistical confidence level.

The combination of the two VLBI points with the $2\sigma$ upper limit at 
144\,MHz from \citet{2026A&A...709L...1B} shows that the spectrum 
of the compact component remains consistent with a single power 
law across the 0.14--5\,GHz range (particularly for $\alpha \gtrsim 0$). 
Consequently, a low-frequency turn-over is not required when only 
the milliarcsecond-scale emission is considered. Moreover, we note 
that \citet{2026A&A...709L...1B} adopted a $2\sigma$ threshold for the 
LOFAR upper limit: a more conventional $3\sigma$ limit would make 
a low-frequency turn-over even less probable when combined with 
the VLBI data. However, this conclusion is largely sensitivity-limited 
given the relatively large uncertainties of the current VLBI data; 
only more sensitive low-frequency VLBI observations can provide 
a definitive comparison with the turn-over reported by \citet{2026A&A...709L...1B}.
This makes FRB\,20190417A only the second PRS, after FRB\,20121102A \citep{marcote17}, for which the radio spectral slope is constrained using measurements obtained entirely at milliarcsecond resolution with VLBI.


\begin{figure*}
\sidecaption
    \includegraphics[width=12cm]{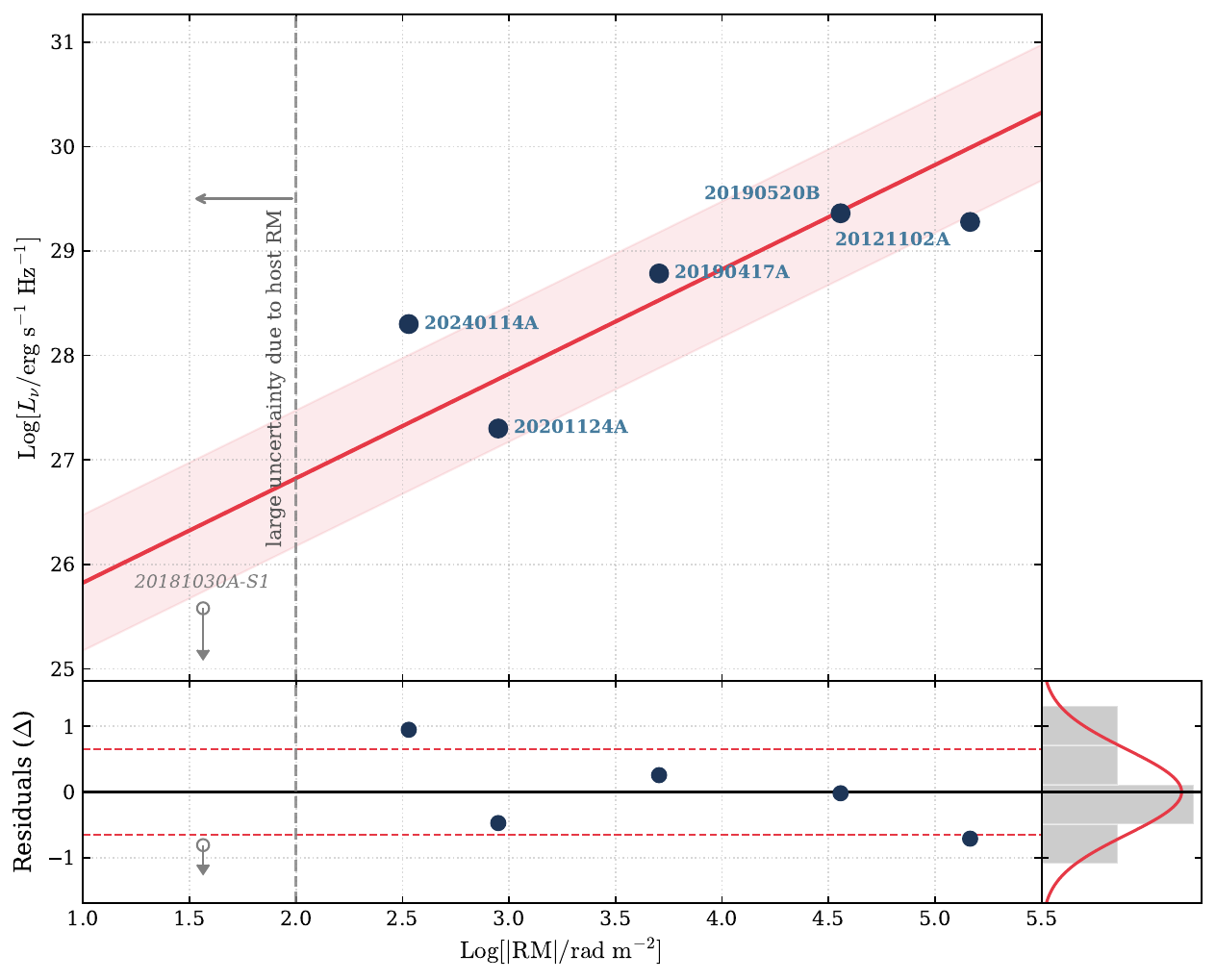}   
    \caption{ $L_\nu$--|RM| relation for five confirmed PRSs and one candidate. Top: Confirmed sources (black circles, labelled), and the candidate 20181030A-S1 \citep{Ibik2024} is shown as a grey point. The solid red line indicates the best-fit relation with a fixed unit slope, and the shaded region shows its $1\sigma$ scatter ($\sigma_\Delta = 0.67$). Bottom: Logarithmic residuals ($\Delta$; left) with $\pm1\sigma$ limits (dashed lines) and their distribution (right) with a Gaussian fit (solid red). Sources with $|{\rm RM}| \lesssim 100~{\rm rad~m^{-2}}$ (left of the dashed green line) may deviate due to contributions of the host ISM.
    }  
    \label{relation}
\end{figure*}
\section{Discussion}

\subsection{Nature of the persistent emission}

The measured spectral index of $\alpha = -0.19 \pm 0.29$ indicates that the spectrum of the PRS associated with FRB 20190417A is nearly flat. When synchrotron radiation is assumed as the most plausible emission mechanism for PRSs, there are two possible interpretations of the observed spectral index. First, the spectrum between 1.5 and 5 GHz might represent the integrated emission from an electron population with a broad energy distribution, characterised by a power-law index of $p = 1 - 2\alpha = 1.4 \pm 0.6$. 
While the nominal value suggests $p < 2$, we note that $p \sim 2$ remains consistent within the $(1-2)\,\sigma$ uncertainty range. Therefore, standard shock acceleration, which typically yields $p \sim 2-3$, cannot be ruled out. This is further supported by the spectral index reported by \citet{Ibik2024}, which, with a smaller uncertainty, is consistent with $p \sim 2-3$. On the other hand, the relatively flat spectrum in our measurement is also compatible with a population of fossil electrons, which often have $p \sim 1-1.5$, as expected in the bubble of a pulsar wind nebula. 
Alternatively, the observed flux spectrum might correspond to the peak of a synchrotron spectrum. 
In this scenario, the peak frequency corresponds to the typical frequency $\nu_m$ related to the minimum Lorentz factor $\gamma_m$ or the synchrotron self-absorption frequency $\nu_a$.
For the former scenario, we have
\begin{align} 
\nu_m\simeq\frac{\gamma_m^2eB}{2\pi m_ec}\sim\nu_{\rm peak}\sim1.5~{\rm GHz},
\end{align}
where $\nu_{\rm peak}$ is the observed peak frequency that is likely at $\sim1.5$ GHz, and $B$ is the magnetic field strength in the emission region. Then, we can obtain the following constraints:
\begin{align}
\left(\frac{\gamma_m}{10^3}\right)^2\left(\frac{B}{1~{\rm mG}}\right)\sim0.54.
\end{align}
If the peak frequency  corresponds to the synchrotron self-absorption frequency $\nu_a$, we have \citep{yan16,Bruni2025AandA}
\begin{align}
\nu_a=\nu_B\left[\frac{\pi e(p-1)n_{e,0}\gamma_m^{p-1}R}{2B}f_{\alpha}(p)\right]^{2/(p+4)}\sim\nu_{\rm peak}\sim1.5~{\rm GHz},
\end{align}
where $n_{e,0}$ is the total electron number density, $\nu_B=eB/2\pi m_ec$ is the electron cyclotron frequency under a magnetic field $B$, and $R$ is the radius of the nebula,  $f_{\alpha}(p)\equiv3^{(p+1)/2}\Gamma[(3p+2)/12]\Gamma[(3p+22)/12]$. 
We assumed $p\sim2$ for a typical particle acceleration mechanism in a shock and obtained the following constraint:
\begin{align}
\left(\frac{B}{1~{\rm mG}}\right)^{2/3}\left(\frac{n_{e,0}}{10^3~{\rm cm^{-3}}}\right)^{1/3}\left(\frac{R}{10^{16}~{\rm cm}}\right)^{1/3}\left(\frac{\gamma_m}{10^3}\right)^{1/3}\sim1.3.
\end{align}

The origin of the nebula that causes PRS and RM remains unsettled, with several viable interpretations. An FRB embedded in a self-absorbed synchrotron nebula can reshape the electron spectrum and generate a spectral hump near the absorption frequency \citep{yan16,Li20}. Alternatively, the PRS might arise from a young magnetar wind nebula powered by synchrotron emission through interactions with supernova ejecta \citep{Murase16,metzger17,margalit18,Rahaman25} or the interstellar medium \citep{Dai17,Yang19}. Other possibilities include a hypernebula driven by super-Eddington outflows in compact binaries \citep{Sridhar22} and accreting wandering massive black holes in dwarf galaxies \citep{Eftekhari20,Reines20,Dong24}.

\subsection{The $L_{\rm radio}$--|RM| relation}
While the precise origin has yet to be determined, a physical connection between the burst source and the PRS can be generally inferred from the concurrent requirements for particle acceleration and Faraday rotation within a magnetised medium. In this context, \citet{yang20,yang22} proposed that the RM of a repeating FRB and its accompanying PRS might originate in a common physical region, leading to a straightforward and nearly model-agnostic connection between the FRB RM and the PRS luminosity.
Building upon this framework, \citet{Yang26} further developed a method that interprets the intrinsic scatter in the $L_\nu - |{\rm RM}|$ relation as a probe of nebular physics. This scatter reflects the growth history of the nebula, parametrised as $R \propto t^{\hat\alpha}$. Using a general scaling $L_\nu \propto R^\epsilon |\mathrm{RM}|$ and examining residuals from the FRB-PRS sample, we can infer the combination of the evolutionary index of $\hat\alpha|\epsilon|$. The latter represents the product of the nebular expansion index ($R \propto t^{\hat{\alpha}}$) and the scaling of the radio luminosity with size ($L_\nu \propto R^{\epsilon}$), offering a robust approach to distinguish between competing models of nebular evolution. Recent studies have also explored the use of the $L_\nu$--|{\rm RM}| relation as a potential standardisable candle for cosmological applications, although its current constraining power is limited by the small sample size, intrinsic scatter, and remaining systematic uncertainties \citep{2025ApJ...984L..40Z,2025ApJ...994..239G}.

Because all currently confirmed PRSs have measurements in the 5–6 GHz band, we adopted the flux densities listed in Table 2 of \citet{Yang26}. For FRB 20190417A, we updated the newly reported flux in this work, $F_\nu = 150~{\rm \mu Jy}$ at 5 GHz. In addition, we included the upper limit of the flux density of the PRS candidate 20181030A-S1 in Fig. \ref{relation}, but excluded it from the calculation of the $L_\nu-|{\rm RM}|$ relation and its scatter.
Following the approach of \citet{Yang26}, we estimated the standard deviation of the residuals using the five currently confirmed PRSs. The measurement uncertainties are negligible compared to the intrinsic scatter and were therefore ignored. We first took the base-10 logarithm of $L_\nu$ and $|{\rm RM}|$ and fitted a linear relation of the form $\log L_{\nu,\rm fit} = \log |{\rm RM}| + C_0$, where the mathematical symbol ``log'' refers to the logarithm to base 10, and $C_0$ represents the mean offset. 
We note that the unit slope was adopted here not as a statistical 
best fit, but as a physically motivated detrending procedure 
(see Sect.~3 of \citealt{Yang26} for details). Given the relation $L_\nu \propto R^\epsilon |{\rm RM}|$, enforcing a slope of unity effectively removes the explicit $|{\rm RM}|$ dependence and isolates the contribution from the stochastic variable $R$ in the residuals.
Importantly, this approach does not require $|{\rm RM}|$ and $R$ to be independent. Any intrinsic coupling between them would manifest as a systematic trend in the residuals, but does not affect the global dispersion. In this case, the resulting scatter provides a robust measure of the dynamic range of $R$ without introducing bias from the fitting procedure.
The residuals were then defined as $\Delta = \log L_\nu - \log L_{\nu,\rm fit}$. The resulting standard deviation of $\Delta$, $\sigma_\Delta = 0.67$, corresponds to $\hat\alpha|\epsilon| \sim 1.5 $. Since FRB 20190417A exhibits a flat spectrum across the observed bands, our results remain consistent with those of \citet{Yang26}. This result is roughly more consistent with scenarios involving forward shocks in the free-expansion phase of SNR/ISM and PWN/SNR systems ($\hat\alpha|\epsilon| \sim 2.0$–$2.8$) and also with young PWNe powered by a nearly steady wind ($\hat\alpha|\epsilon| \sim 1$).


\section{Conclusions}

In this Letter, we reported on observations made with the European VLBI Network (EVN) at 5 and 8 GHz of the persistent radio source associated with FRB\,20190417A, and we provided upper limits for the candidate PRS associated with FRB\,20181030A. Our main results are summarised below.

\begin{itemize}

\item At 5 GHz, we detected a compact source at milliarcsecond scales in a location that is consistent with the PRS linked with FRB\,20190417A by previous studies \citep{2026ApJ...996L..16M}. The source is unresolved, with a projected physical size \(\lesssim 15\) pc, and a high brightness temperature ($T_{\rm b} > 3.8 \times 10^{5}$ K), supporting a non-thermal synchrotron origin. At 8 GHz, we derived a constraining upper limit on its emission.

\item By combining our measurement with VLBI observations at 1.4\,GHz, we derived a radio spectral index $\alpha = -0.19 \pm 0.29$. This makes FRB\,20190417A only the second PRS, after FRB\,20121102A, with a spectral index constrained entirely using VLBI data. The measured value indicates a relatively flat spectrum, corresponding to an electron power-law index $p = 1-2\alpha = 1.4 \pm 0.6$. While the nominal value favours a hard spectrum ($p<2$), the conventional value $p\approx2$ (as expected from diffusive shock acceleration) remains consistent within the $\sim$1$\sigma$ uncertainty range. The nearly flat spectrum can also be interpreted as the peak of a synchrotron spectrum, providing constraints on the physical conditions of the emitting region (e.g.\ magnetic field, particle density, and size).

\item We placed FRB\,20190417A in the $L_\nu$--|RM| plane, where it is consistent with the proposed relation linking PRS luminosity and Faraday rotation measure. When we included this source, we estimated a scatter of $\sigma_\Delta = 0.65$, corresponding to $\hat{\alpha}|\epsilon| = 1.5 \pm 0.7$, which is consistent with scenarios involving young pulsar wind nebulae or forward shocks in the free-expansion phase.

\item For the candidate PRS 20181030A-S1, we reported non-detections at 5 and 8 GHz. At 5 GHz, the upper limit implies a spectral luminosity $L_{5\,\mathrm{GHz}} \lesssim 3.8\times10^{25}\ {\rm erg\ s^{-1}\ Hz^{-1}}$ and constrains the spectral index to $\alpha \lesssim -1.2$ relative to the VLA measurement at 1.5 GHz. This would indicate an unusually steep spectrum if the emission were to arise from a compact source. Alternatively, the VLA detection might be dominated by diffuse host-galaxy emission (e.g. star-forming regions) and not directly associated with the FRB.

\end{itemize}
\begin{acknowledgements}
We thank A. Moroianu for helping us compare the results from the PRECISE collaboration with those presented in this work.
Y.P.Y is supported by the National Natural Science Foundation of China (No. 12473047), the National Key Research and Development Program of China (No. 2024YFA1611603) and the Yunnan Key Laboratory of Survey Science (No. 202449CE340002).
The research leading to these results has received funding from the European Union’s Horizon 2020 programme under the AHEAD2020 project (grant agreement no. 871158).
The European VLBI Network is a joint facility of independent European, African, Asian, and North American radio astronomy institutes. Scientific results from data presented in this publication are derived from the following EVN project codes: EB116A and EB116B. 

\end{acknowledgements}

   \bibliographystyle{aa} 
   \bibliography{Main} 
%
\appendix
\section{EVN observations and data processing}


Observations were performed with the EVN in two sessions, EB116A (May 30, 2025) and EB116B (June 18, 2025), at central frequencies of 8 (X-band) and 5 GHz (C-band), respectively. The observations were conducted in phase-referencing mode to enable accurate astrometry and high sensitivity to faint compact emission. The target source 20181030A-S1 was calibrated using the phase-reference source J1048+7143. The target source 20190417A-S1 was calibrated using the phase-reference source J1934+6138. A phase referencing cycle of 
$\sim$5 minutes was scheduled for both bands. These were dedicated continuum observations aimed at detecting the persistent radio sources only. No dedicated real-time or offline search for bursts was performed on these data.
However, no bursts from either FRB\,20190417A or FRB\,20181030A were detected with other telescopes during the EVN observations.

The EB116A experiment was carried out with five antennas: Westerbork (Wb), Effelsberg (Ef), Onsala (O6), Toru\'n (Tr), and Irbene (Ib). 
Nine antennas participated in EB116B: Jodrell Bank (Jb), Westerbork (Wb), Effelsberg (Ef), Onsala (O8), Tianma (T6), Urumqi (Ur), Toru\'n (Tr), Irbene (Ib), and Sardinia Radio Telescope (Sr). 
In both sessions, data were recorded in dual circular polarisation. 
The data were correlated at the Joint Institute for VLBI ERIC (JIVE) using the SFXC correlator with 1-second integrations, producing eight 32-MHz subbands with 64 spectral channels.

The data were calibrated following standard EVN procedures within the 
Astronomical Image Processing System ({\tt AIPS}, \citealt{greisen2003}). 
This included a priori amplitude calibration using system temperatures 
and gain curves ({\tt ANTAB} and {\tt APCAL} tasks), parallactic angle correction, 
and global fringe fitting on nearby phase calibrators. 
No amplitude self-calibration was performed, as the target sources are 
too faint for reliable self-calibration. The amplitudes of the phase calibrators were verified 
to agree with their catalog values within a few percent, confirming the 
reliability of the a-priori flux scale (typical uncertainty $\sim$10\%).
The calibrated visibilities were imaged in {\tt Difmap} \citep{1997ASPC..125...77S} using natural weighting (to minimize image noise) and the {\tt CLEAN} algorithm. Due to the low signal to noise ratio ($\sim10$), and to preserve the astrometric information of the target, no amplitude or phase self-calibration were applied. The resulting clean images were then imported 
into {\tt CARTA} \citep{2026PASP..138b4506W}, where an elliptical Gaussian component was fitted in the 
image plane to measure the position and integrated flux density of the 
source. In case of non-detection, the RMS noise was measured in a region centred on the expected PRS position from \cite{Ibik2024}.

Table \ref{tab:evn_obs} summarizes the EVN observations and results.


\begin{table*}
\centering
\caption{EVN observations and main results for the two candidate persistent radio sources.}
\label{tab:evn_obs}
\begin{tabular}{lcccc}
\hline
\hline
Target & Frequency & Beam size & RMS & Integrated flux density \\
       & (GHz)     & (mas)     & ($\mu$Jy beam$^{-1}$) & ($\mu$Jy) \\
\hline
\multirow{2}{*}{20190417A-S1} 
       & 5.0       & $6.5 \times 3.4$ & 16 & $150 \pm 45$ \\
       & 8.0       & $6.6 \times 3.5$ & 50 & $<250$  \\
\hline
\multirow{2}{*}{20181030A-S1} 
       & 5.0       & $6.5 \times 3.4$ & 16 & $<80$  \\
       & 8.0       & $5.5 \times 3.6$ & 30 & $<150$  \\
\hline
\end{tabular}
\tablefoot{All upper limits are quoted at the 5$\sigma$ level. The source 20190417A-S1 is unresolved at 5\,GHz.}
\end{table*}


\section{Detection and localisation of 20190417A-S1}
\label{appendix:detection}
\begin{figure}
    \centering
    \includegraphics[width=0.5\textwidth]{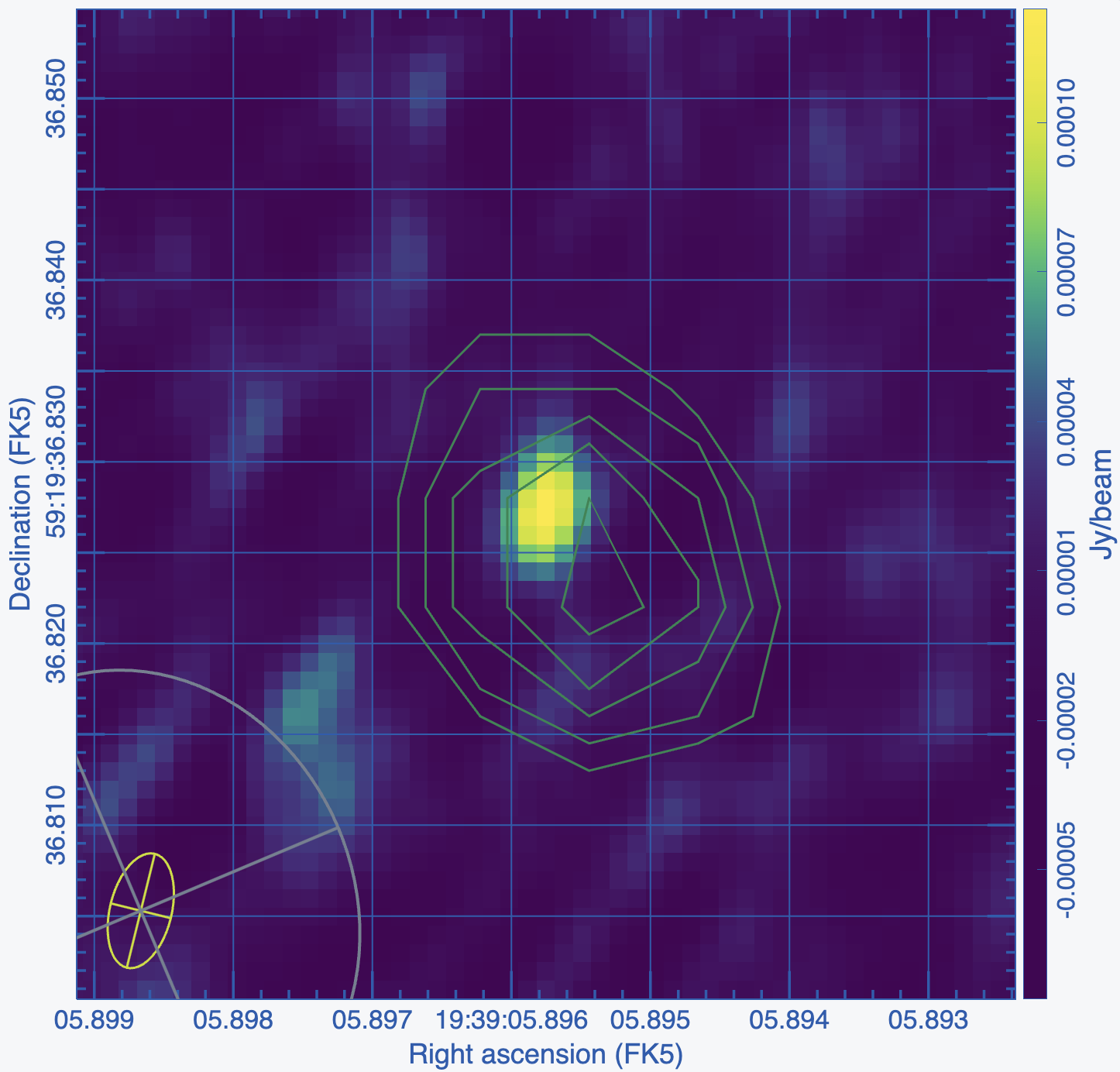}
    \caption{EVN 5\,GHz image of the persistent radio source associated with FRB\,20190417A (colour scale). 
Contours show the EVN 1.4\,GHz detection from \cite{2026ApJ...996L..16M}, drawn at levels of 
5, 6, 7, 8, 9 and 10 times the RMS noise (21 $\mu$Jy\,beam$^{-1}$). 
The synthesised beams are indicated in the lower left corner (grey ellipse for 1.4\,GHz and 
yellow ellipse for 5\,GHz).}
    \label{fig:PRS}
\end{figure}

In our EVN 5 GHz observations, we detect a compact radio source at a position consistent with 20190417A-S1. The synthesised beam (FWHM) is $6.5 \times 3.4$ mas with a position angle of $-13.8^\circ$, and the image RMS noise is $16\ \mu$Jy\,beam$^{-1}$.
The source is well described by a single Gaussian component. The best-fit position (referenced to the FK5 system) is
\begin{align*}
\alpha_{\rm PRS}~({\rm J2000}) &= 19^{\rm h}39^{\rm m}05.89575^{\rm s} \pm 0.7~{\rm mas}, \\
\delta_{\rm PRS}~({\rm J2000}) &= +59^\circ19'36.8276'' \pm 0.7~{\rm mas}.
\end{align*}
The corresponding ICRS coordinates are 
$\alpha = 19^{\rm h}39^{\rm m}05.8942^{\rm s} \pm 0.7~{\rm mas}$ and 
$\delta = +59^\circ19'36.8057'' \pm 0.7~{\rm mas}$.
To account for systematic uncertainties, the quoted positional error of $\sim$0.7 mas in both coordinates is the quadratic sum of (i) the formal Gaussian-fit uncertainty, (ii) 10\% of the synthesised-beam major axis (to conservatively include residual phase-referencing and atmospheric contributions), and (iii) the absolute positional uncertainty of the phase calibrator. After calibration, the measured position of the phase calibrator agrees with its catalog position to within 0.1 mas, confirming that phase-referencing errors are negligible. The location of our counterpart is consistent within 3-$\sigma$ with that of \citet[considering the larger uncertainties of their 1.4 GHz measurement of $\sim$4 mas]{2026ApJ...996L..16M}. In Fig.~\ref{fig:PRS} we show both detections, with the 1.4 GHz EVN image from \citet[reproduced from archival data, project EK050G]{2026ApJ...996L..16M} displayed in contours.

The source has an integrated flux density of $S_{\rm 5\,GHz} = 150 \pm 45\ \mu$Jy, as extracted via Gaussian fitting. The given uncertainty includes both the Gaussian fit error and a 10\% uncertainty on the absolute flux density scale, added in quadrature.
The fitted source size is $\theta_{\rm maj} = 6.4 \pm 1.4$ mas and $\theta_{\rm min} = 3.8 \pm 0.5$ mas, consistent with the synthesised beam. We therefore consider the source as unresolved at milliarcsecond scales. Adopting the beam major axis as the most conservative upper limit on the angular size, this implies a projected physical size \(\lesssim 15\) pc at the redshift of the host galaxy (\(z=0.12817\), \citealt{Ibik2024}), and a brightness temperature $T_{\rm b} > 3.8 \times 10^{5}$ K, supporting a non-thermal synchrotron origin for the emission. Finally, the integrated flux density measured with the EVN corresponds to a spectral luminosity of $L_{5\,\mathrm{GHz}} = (6.2 \pm 1.9)\times10^{28}\ {\rm erg\ s^{-1}\ Hz^{-1}}$, consistent within errors with the one reported by \cite{2026ApJ...996L..16M}. At 8 GHz, the source is not detected, resulting in a 5-$\sigma$ upper limit of 250 $\mu$Jy.


\section{Upper limits on 20181030A-S1}
\label{appendix:nondetection}
For the candidate PRS 20181030A-S1 we obtained a non-detection both at 5 and 8 GHz. Assuming the host distance of $\sim 20$ Mpc for FRB\,20181030A -- as it was associated with NGC\,3252, see \cite{Bhardwaj_2021} -- our $5$ GHz upper limit of $80\,\mu$Jy (5-$\sigma$) corresponds to a spectral luminosity upper limit of $L_{5\,\mathrm{GHz}} \lesssim 3.8\times10^{25}\ {\rm erg\ s^{-1}\ Hz^{-1}}$.
%
%
At 8 GHz, the non detection is less stringent, with a 5-$\sigma$ upper limit of $150\,\mu$Jy.

At 5 GHz, the non-detection places a strong constraint on the radio spectral index, implying a very steep spectrum ($\alpha \lesssim -1.2$) when compared with the VLA flux density measured at 1.5 GHz by \cite{Ibik2024}. Alternatively, the emission detected at VLA resolution may be dominated by diffuse components within the host galaxy (e.g. star-forming regions), rather than being directly associated with the FRB.


\end{document}